# Superconducting detector of IR single-photons based on thin WSi films

V A Seleznev[1,2], A V Divochiy[1,2], Yu B Vakhtomin[1,2], P V Morozov[2], P I Zolotov[1,2], D D Vasil'ev[3], K M Moiseev[3], E I Malevannaya[3] and K V Smirnov[1,2,4]

[1] Moscow State Pedagogical University, 119435 Moscow, Russia
[2] CJSC "Superconducting nanotechnology" (Scontel), 119021 Moscow, Russia
[3] Bauman Moscow State Technical University, 105005 Moscow, Russia
[4] National Research University Higher School of Economics, 101000 Moscow, Russia

E-mail: seleznev@rplab.ru

**Abstract**. We have developed the deposition technology of WSi thin films 4 to 9 nm thick with high temperature values of superconducting transition ($T_c$~4 K). Based on deposed films there were produced nanostructures with indicative planar sizes ~100 nm, and the research revealed that even on nanoscale the films possess of high critical temperature values of the superconducting transition ($T_c$~3.3-3.7 K) which certifies high quality and homogeneity of the films created. The first experiments on creating superconducting single-photon detectors showed that the detectors' SDE (system detection efficiency) with increasing bias current ($I_b$) reaches a constant value of ~30% (for $\lambda$=1.55 micron) defined by infrared radiation absorption by the superconducting structure. To enhance radiation absorption by the superconductor there were created detectors with cavity structures which demonstrated a practically constant value of quantum efficiency >65% for bias currents $I_b \geq 0.6 \cdot I_c$. The minimal dark counts level (DC) made 1 s$^{-1}$ limited with background noise. Hence WSi is the most promising material for creating single-photon detectors with record SDE/DC ratio and noise equivalent power (NEP).

## 1. Introduction
Development of thin films superconducting technologies has created multiple superconducting low-current electronics devices among which there can be marked out a group of electro-magnetic radiation sensors: mixer based on SIS Josephson junction (superconductor-insulator-superconductor)[1], direct bolometers and heterodyne receivers based on hot electron phenomena in superconductors (HEB, Hot Electron Bolometer)[2], bolometers based on superconducting transition (TES, transition edge sensor)[3], detectors based on kinetic inductance (KID, kinetic inductance detector)[4], superconducting single photon detector (SSPD)[5] etc. Progress in developing different detectors is as a rule ensured by development of the technologies of creating structures and by application of new superconducting materials. For example the initially discovered effect of superconductivity destruction at absorption of single infrared photon by a thin and narrow NbN strip biased by current close to critical value has allowed creating detectors of single infrared photons with quantum efficiency, or photons detection probability, at the level of several percent [5],[6]. Whilst usage of other superconducting materials such as NbTiN [7], MoRe [8], WSi [9], MoSi [10], NbSi [11], TaN [11], and improve of the technology of creating superconducting single photon detectors using resonator structures and antireflection coatings allowed creating a superconducting single photon detector with increased quantum efficiency value up to several dozen percent [13] and improved values of all main characteristics of the detectors: counting rate [14], Dark Count probability [15], timing resolution [16].

One of the problems in creating SSPD is to get detectors with simultaneous achievement of high level of detection efficiency and low level of dark counts. Both SDE and DC depend on bias current $I_b$, both indexes go down while displacement current goes down. As a rule, DC reduction along with $I_b$ reduction is linked to fluctuation processes reductions (temperature, normal electrons and Cooper pairs concentricity) [17],[18]. While SDE reduction along with the detector's bias current reduction is linked to inhomogeneity of the superconducting strip (thickness, width, composition and structure of film) and, as a rule, difficulties in achieving the internal detection efficiency close to 100%; the situation when each photon absorbed by the superconducting strip leads to destruction of superconductivity across the whole cros-section of the superconducting strip. Creation of highly homogeneous superconducting films and films-based structures is one of the most important tasks of producing SSPD with the lowest NEP or a high SDE/DC ratio. We have to note that the records for different types of SSPD published until now show considerably falls of SDE when the bias current goes down.

This study is aimed to elaborate and create superconducting single photon detectors with saturation of current dependence of $SDE(I_b)$ at the currents range of $0.6 \cdot I_c < Ib < I_c$. When choosing materials for the detectors we first took into account experience of creating SSPD with high SDE values. When comparing with NbN (or NbTiN as close to it) we chose WSi for several reasons. First of all, NbN films are polycrystalline with axial structure aligned with the film's thickness. At that, the size of crystallites depends much on the parameters of the film deposition and on the substrate type [19]. Achieving the highest homogeneity level of the NbN film on considerable areas comparable to typical SSPD active area is absolutely not a trivial task. In the meantime as testified by several research groups [20] WSi has amorphous structure which makes the aim of creating homogeneous superconducting nanostructures still more technological. Besides, WSi have a lower value of energy gap that also assumes possibility of producing detectors with highest values of internal DE. Hence in this study we present the results of elaboration of the technology of WSi thin films deposition and creation of superconducting single photon detectors for the near infrared range with internal detection efficiency close to 100% for bias current range of $0.6 \cdot I_c < I_b < I_c$, with high potential to realize record-breaking values of NEP and SDE/DC ratio.

**2. Deposition of WSi thin films**
For deposition tungsten silicide film we used the magnetron sputtering method realized on the basis of VUP-11M technological unit. W and Si were sputtered simultaneously from two magnetron sources. To sputter Si there was used a magnetron working in high frequency mode. To sputter W there was used a magnetron working in impulse mode at 15 kHz frequency. Magnetrons were located at the angle of 90° to each other and the substrate holder at the angle of 45° to axial lines of both magnetrons. As planar sizes of the substrate in our experiments did not exceed 20·20 mm$^2$ and the distance between each magnetron center and the substrate exceeded 15 cm, our method allowed insuring high level of the film homogeneity [21], confirmed by measurements of sheet resistance and temperature of transition into superconducting state taken at different areas of the substrate. Deviations of the above mentioned values within the substrates did not exceed 5%. We must also mention that the method employed and the technique realized allows to easily changing the ratio of W and Si materials in produced films by varying magnetron power. This fact is doubtless importance while optimizing the process of films deposition.

While optimization of ratio of W and Si as a criterion the film's optimal composition we used its critical temperature value for superconducting transition ($T_c$). These researches were made for the film thickness fixed at 4 nm. Normalized dependencies for films resistance on temperature are shown on figure 1. As a substrate we used high-resistivity Si-wafer of 400-micron thick with a thermal silicone dioxide layer of 250-nm thick. The chart we have obtained shows that when percentage of W in WSi films is changed within the range of 71÷81% the critical temperature of the film's changed within the range of 2.78 K to 3.35 K. The highest $T_c$ value was achieved with atomic ratio $W_{0.76}Si_{0.24}$. It can be also noted that dependencies of R(T) for WSi films had a typical 'metallic like' rate marked by reduction of resistance upon changes in temperature from 300K to 4K . Electrical resistance of the films made about 200·μOhm·cm did not practically depend on W content within 71-81%, and complied with the data published earlier [9],[20],[21]. The parameters of WSi films deposition process corresponding to the optimal atomic ratio are shown in table 1. In further work we used WSi films of 5 nm, 7 nm and 9 nm thick with the highest $T_c$ value corresponding to the atomic ratio $W_{0.76}Si_{0.24}$ which made $T_c$(5 nm)=3.65 K, $T_c$(7 nm)=3.75 K и $T_c$(9 nm)=3.8 K.

**Table 1.** Parameters of WSi films deposition process.

| | |
|---|---|
| P (residual), mbar | $3.0·10^{-5}$ |
| P (operating), mbar | $2.4·10^{-3}$ |
| Argon flow, l/h | 2.6 |
| W magnetron power, watt | 120 |
| Si magnetron power, watt | 109 |
| W deposition rate, nm/sec | 0.30 |
| Si deposition rate, nm/sec | 0.08 |

### 3. WSi single photon detectors and measuring methods
*3.1. Technology WSi SSPD fabrication.*

Using $W_{0.76}Si_{0.24}$ films 5, 7, 9-nm thick we produced single photon detectors which was in shape of a superconducting strip ~120 nm wide bended into meander with ~200 nm step and covering the area of 7·7 micron$^2$ (active area of the detector). The superconducting strip had contact pads set as a coplanar line. The technology of creating the detectors was based on direct e-beam lithography and following reactive plasma etching and was similar to NbN detectors technology [22]

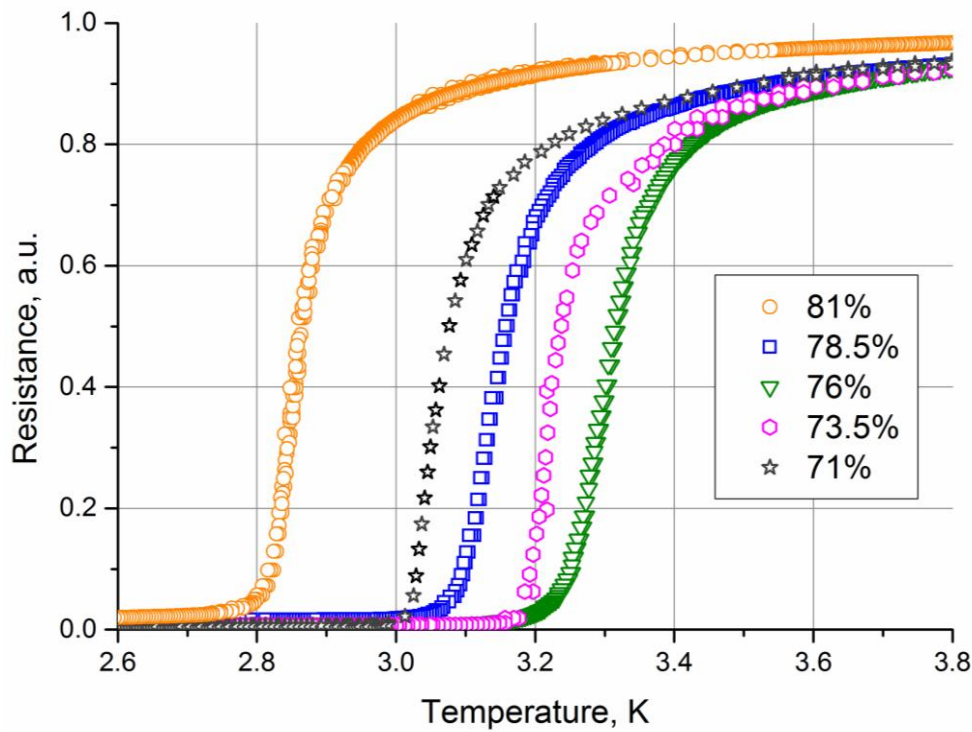

**Figure 1.** Temperature dependencies of resistance of $W_xSi_{1-x}$ 4 nm thick films. x-values percentage shown on the insert to the Figure.

We produced WSi detectors on high-resistivity silicon wafers 400 microns thick with a thermal silicone dioxide layer 250-nm thick (detector type I). For the 2$^{nd}$ type of detector we apply a resonator structure before deposition of a WSi film. The resonator structure consists of a metal mirror and a quarter-wave dielectric layer (optimized for λ=1.55 micron). The metal mirror of 50·50 micron$^2$ was made of a 70-nm thick gold layer applied with e-beam evaporation method by using direct photo lithography and following wet etching. The metal mirror was located under the SSPD active area. A silicon nitride layer 180-nm thick applied by PECVD method onto the whole substrate surface acted as quarter-wave resonator layer (figure 2). Applying a resonator structure we solved the problem of increasing radiation absorption by a superconductor that does not exceeds 30% in case of simple Si/SiO$_2$ substrate.

*3.2. Coupling the SSPD with radiation.*

In our experiments we used direct coupling of the WSi detector with radiation through standard single-mode fiber Corning SMF 28e. Since the active area of the detector made 7·7 micron$^2$, mode field diameter for that fiber type at λ=1.55 micron made ~9 micron and in-fiber radiation have a Gaussian profile, then the best geometrical coupling in our experiments was ~85%.

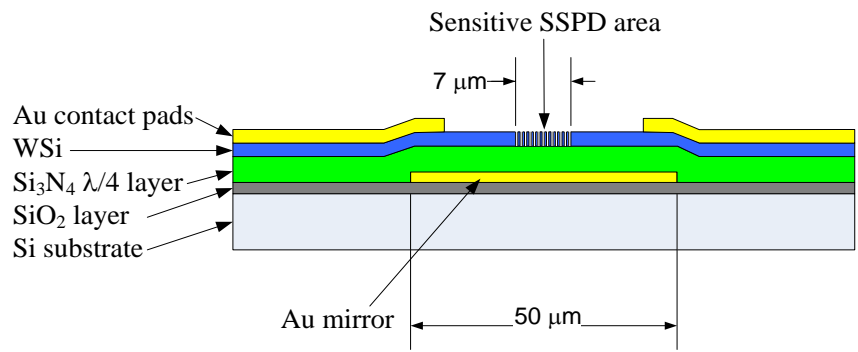

**Figure 2.** Schematic the WSi detector cross-section having resonator structure.

To accurate positioning the detector against the fiber we used two methods. The first one is based on applying a flip-chip positioning device Fineplacer Lambda which allows observing and positioning two space-divided objects against each other (the fiber core and the detector's active area in our case), and then automatically space-match them with accuracy no worse than 0,5 micron. The second method is based on possibility to observe the bolometric signal from the superconducting detector at room temperature at absorption of modulated optical radiation [14]. By bringing the optical radiation via single-mode fiber and under the condition the detector is biased by DC current it becomes possible to define how accurately the fiber is positioned against SSPD. After reaching the maximum signal the fiber was fixed against the detector.

*3.3. Low-temperature technique of the experiment.*

The experiments for defining WSi films superconducting transition temperature and measuring SDE and DC of WSi single photon detectors were made with a low-temperature technique that allowed temperatures down to 1.6K. The technique is based on dipstick into a standard liquid helium Dewar. Helium entered to the dipstick internal volume through capillary that connected to the helium Dewar bath. The double-wall dipstick with vacuum between the walls allowed for thermal insulation of the dipstick's internal volume from surrounding. Evacuation of the helium vapor from the dipstick's internal volume with a forevacuum pump to reach the pressure of ~10 mbar allowed reducing the temperature in the dipstick down to 1.6K.

*3.4. SDE and DC measuring methods.*

The block chart of the set-up we applied for measuring SDE and DC detectors is shown on figure 3. As a source of photons for detection efficiency measuring we employed a fiber laser diode with 1550 nm wavelength in a continuous wave generation mode. Radiation was going from the source via a calibrated fiber coupler into two channels: detector channel and powermeter channel. Radiation power was measured by Ophir IRG300 powermeter with a accuracy better than 1 picowatts. In the detector channel radiation was passing through a calibrated sensitive attenuator and further through the single-mode fiber to the detector. The radiation power level was fixed at $10^7$ photons per second coming to the single-mode fiber connected to the WSi single photon detector. Voltage pulses from the detector passed through the bias-T adapter into the amplification chain consisting of two ultra wideband Mini-circuit amplifiers (bandwidth 0.1-1000 MHz, total gain ~46 dB), and further to the Agilent 53131A pulse counter. To control the shape and scape of the pulse the signal was also send to oscilloscope. The bias-T adapter, an amplification chain and DC biasing unit were integrated into a single unit.

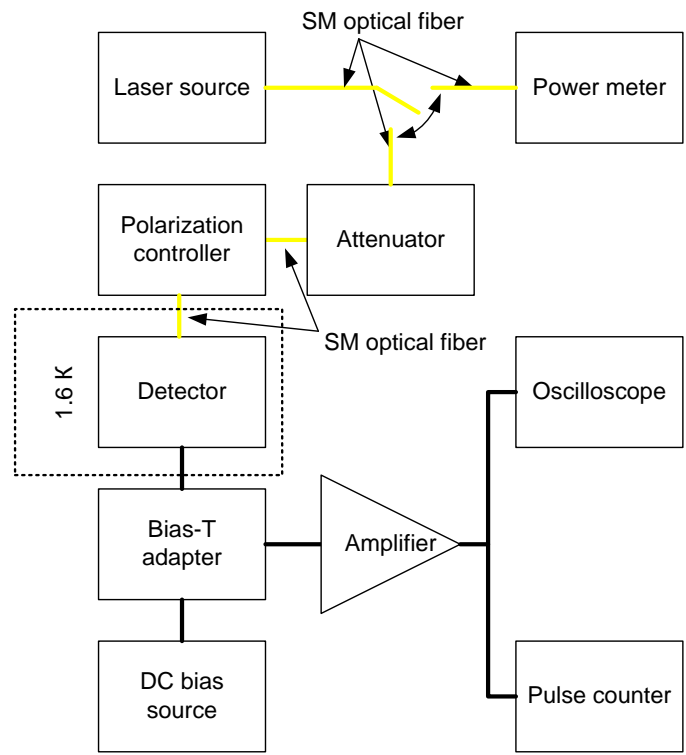

**Figure 3.** Block chart of the set-up for measurement SDE and DC of single photon detectors based on WSi films.

Detection efficiency of the detector was defined as a ratio between number of electrical pulses coming from the detector per second to the number of photons reaching the detector via single-mode fiber within the same time period. The detection efficiency defined above is the System Detection efficiency because it does not take into account radiation loss related to optical coupling between single mode optical fiber and detector as well as losses related to a certain radiation absorption by the superconductor.

For measuring the dark count rate we made use of the same experimental set-up but in case of screening from radiation of the standard single-mode optical fiber input connected to the detector. For measuring the lowest dark count rates the integration time of the pulse counter was extended to 10 seconds.

## 4. Results of measurements

To define an optimal thickness of superconducting WSi films in order to reach best SDE/DC ratio and NEP) we produced single-photon detectors based on 5-nm, 7-nm and 9-nm thick films. Figure 4 shows dependencies of SDE and DC for these detectors from relative bias current $I_b/I_c$, with $I_b$ being the detectors current and $I_c$ being the critical current. We must note that at maximum bias currents the single photon detectors based on 5-nm and 7-nm films showed SDE equal to ~20-30% for $\lambda=1.55$ μm, which is determined by the maximum radiation absorption and optical coupling between single-mode optical fiber and detector. Whilst the detector based on the 9-nm film had an absolute value of 15% for the same wavelength. For making it more convenient to compare and analyze dependencies of SDE ($I_b/I_c$) for different detectors the present dependencies normalized to the absorption coefficient. DC values are given in the chart in absolute units. The dependencies obtained certify the following: a) higher levels of SDE for the detectors based on 5-nm and 7-nm thick films are determined by the absorption coefficient but achieved at different $I_b/I_c$ values; SDE values for the detectors based on 9-nm thick film do not reach saturation; b) enduring saturated dependence SDE ($I_b/I_c$) is only indicative for the detectors produced from 5-nm thick WSi films; saturation arise already at currents of ~0.75 $I_b/I_c$; c) DC dependencies ($I_b/I_c$) are similar for detectors based on films of different thickness.

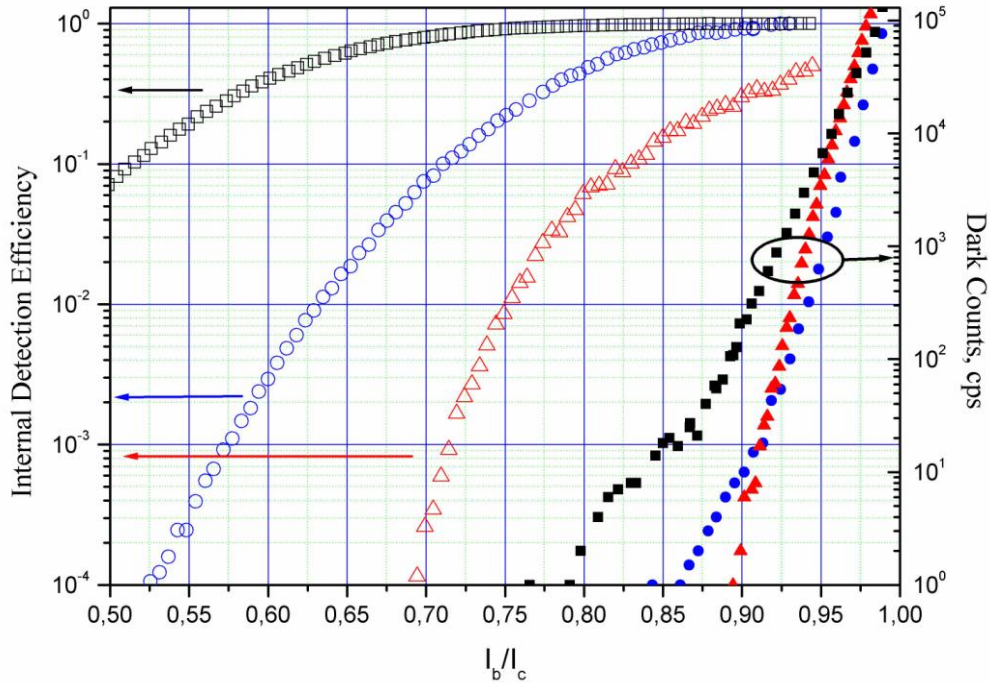

**Figure 4.** Standardized dependence of SDE (open symbols) and DC (closed symbols) for single photon detectors on Si/SiO$_2$ substrate from the modified displacement current for detectors based on WSi films of the following thickness: □, ■ - 5 nm, ○, ● - 7 nm, △, ▲ - 9 nm.

Presented SDE dependencies ($I_b/I_c$) allow for the following conclusions. First of all as the detection efficiency for the 5-nm film detector reaches its constant value at $I_b/I_c>0.75$ and does not grow further together with the detector bias current. It allow to conclude that the internal quantum efficiency of this detector reaches its maximum level of 100% i.e. each photon absorbed by the superconducting structure generates a voltage pulse. Hence we specified the internal detection efficiency on normalized system detection efficiency chart, the first reaching its maximum value of 1 at $I_b/I_c>0.75$. Besides, reduction of the bias current within this currents range causes considerable reduction of the dark count rates down to 1 cps with the quantum efficiency preserved at a constant level.

The detector based on 7-nm thick film also reaches its highest value of the internal quantum efficiency but this takes place solely with maximum levels of bias currents. Despite the 5-nm film detector, this detector reacts onto reduction of dark count rates during the displacement current reduction by reducing considerably its quantum efficiency. The detector based on 9-nm thick film has a still stronger dependence of the detection efficiency on bias current and its internal quantum efficiency does not reach its maximum. We must note that saturated dependence of SDE vs. bias current was not earlier observed for single photon detector of NbN films. We believe that strong saturation of SDE dependence ($I_b/I_c$) is connected to the WSi films deposition and structuring technology when such parameters as thickness and width the superconducting strip, the superconductor composition and stoichiometric formula are homogeneous well. Undoubtedly the main advantage of the created composition is the possibility of reaching maximum values of SDE/DC and NEP by reducing the detector displacement current and correspondingly reducing DC while preserving SDE value as unchanged. It is also evident that highest levels of SDE/DC and NEP can be in demand just in case high absolute level of SDE reached.

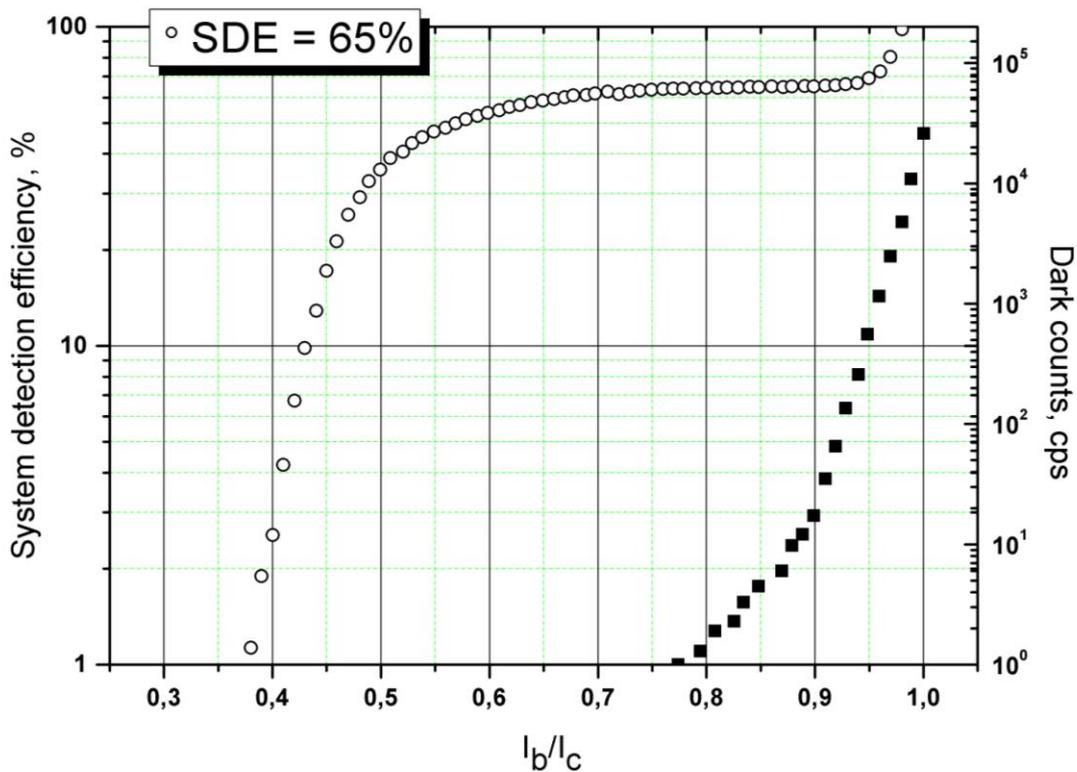

**Figure 5.** Dependence of SDE (open symbols) and DC (closed symbols) for single photon detectors on Si/Au/Si$_3$N$_4$ substrate with $\lambda/4$ resonator layer from the modified displacement current for 5-nm thick WSi films detectors.

To increase SDE of the 5-nm WSi films based structures we created detector with $\lambda/4$ resonator structure located under the superconducting film. Technology and topology of these structures are described above. SDE (Ib/Ic) and DC ($I_b/I_c$) dependencies for such detectors are shown on Figure 5 and have the dependencies equal to detector based Si/SiO$_2$ substrate. The maximum absolute value of SDE measured for such detectors makes 65% for a 1.55 μm wavelength. Constancy of the system quantum efficiency is observed in a wider range of bias currents, that is in the area of >0.6$I_b/I_c$. DE dependence ($I_b/I_c$) of the created detector shows that when the displacement current is reduced the dark count rate falls down to 1 cps with a constant SDE value. Unfortunately with further reduction of the bias current it proved impossible to reduce the DC level because of the background radiation. The additional cooled filters for the background radiation [23] if used can allow for further reduction of DC in the area of $0.6 I_b/I_c$ where the system quantum efficiency is still of constant value.

The best NEP value usually determined for quantum detectors as $NEP = \frac{h\nu}{SDE}\sqrt{2DC}$, where $h\nu$ is photon energy, made $4.4 \cdot 10^{-20}$ W·Hz$^{-1/2}$ (for $\lambda=1.55$ μm, SDE=65 %, DC=1 cps), and the ratio of SDE/DC=0.65 s. Despite the fact that the authors [13] have come to SDE=90% when using WSi single photon detectors, the best SDE/DC ratio obtained in this study made SDE/DC=$3 \cdot 10^{-3}$ s, with the value of NEP=$3.9 \cdot 10^{-19}$ W·Hz$^{-1/2}$, which is considerably worse than the results presented in this study owing to the lack of dependence of SDE ($I_b/I_c$) shows a constant value of the system detection efficiency in a wide range of bias currents.

## 5. Conclusion

Whereas this study presents technology of deposition and structuring of WSi ultrathin films and creating superconducting single photon detector for telecommunication wavelengths ($\lambda=1.55$ μm) which can simultaneously reach high values of the system quantum efficiency (65%) and low levels of the dark count rate (1 cps). Obtained records of NEP=$4.4 \cdot 10^{-20}$ W·Hz$^{-1/2}$ (for $\lambda=1.55$ μm) and SDE/DC=0.65 s are the best amongst superconducting single photon detectors and in comparison with other alternative types of single photon detectors.

## 6. References

[1] Tucker J R and Feldman M J 1985 Quantum detection at millimeter wavelengths *Reviews of Modern Physics* **57(4)** 1055
[2] Shurakov A, Lobanov Y and Goltsman G 2015 Superconducting hot-electron bolometer: from the discovery of hot-electron phenomena to practical applications *Superconductor Science and Technology* **29(2)** 023001
[3] Lita A E, Miller A J and Nam S W 2008 Counting near-infrared single-photons with 95% efficiency *Opt. Express* **16** 3032–304
[4] Doyle S, Mauskopf P, Naylon J, Porch A and Duncombe C 2008 Lumped element kinetic inductance detectors *Journal of Low Temperature Physics* **151(1-2)** 530-536
[5] Gol'tsman G N et al. 2001 Picosecond superconducting single-photon optical detector *Appl. Phys. Lett.* **79** 705–707
[6] Semenov A D, Gol'tsman G N and Korneev A A 2001 Quantum detection by current carrying superconducting film *Physica C* **351** 349–356
[7] Dorenbos S N, Reiger E M, Perinetti U, Zwiller V, Zijlstra T and Klapwijk T M 2008 *Applied Physics Letters* **93** 131101
[8] Gol'tsman G N et al. 2007 Middle-infrared ultrafast superconducting single photon detector *Infrared and Millimeter Waves 15th International Conference on Terahertz Electronics IRMMW* (Hong Kong) 115-116
[9] Baek B, Lita A E, Verma V and Nam S W 2011 Superconducting a-W$_x$Si$_{1-x}$ nanowire single-photon detector with saturated internal quantum efficiency from visible to 1850 nm *Appl. Phys. Lett.* **98** 251105
[10] Korneeva Y P, Mikhailov M Y, Pershin Y P, Manova N N, Divochiy A V, Vakhtomin Y B, and Goltsman G N 2014 Superconducting single-photon detector made of MoSi film *Superconductor Science and Technology* **27** 095012
[11] Dorenbos S N, Forn-Diaz P, Fuse T, Verbruggen A H, Zijlstra T, Klapwijk T M and Zwiller V 2011 Low gap superconducting single photon detectors for infrared sensitivity. *Appl. Phys. Lett.* **98** 251102
[12] Engel K A, Aeschbacher A, Inderbitzin K, Schilling A, Ilin K, Hofherr M, Siegel M, Semenov A and Hubers H-W 2012 Tantalum nitride superconducting single-photon detectors with low cut-off energy *Appl. Phys. Lett.* **100** 062601

[13] Marsili F, Verma V B, Stern J A, Harrington S, Lita A E, Gerrits T and Nam S W 2013 Detecting single infrared photons with 93% system efficiency *Nature Photonics* **7(3)** 210-214
[14] Sidorova M V, Divochiy A V, Vakhtomin Y B and Smirnov K V 2015 Ultrafast superconducting single-photon detector with a reduced active area coupled to a tapered lensed single-mode fiber *Journal of Nanophotonics* **9(1)** 093051-093051
[15] Smirnov K, Vachtomin Y, Divochiy A, Antipov A and Goltsman G 2015 Dependence of dark count rates in superconducting single photon detectors on the filtering effect of standard single mode optical fibers *Applied Physics Express* **8(2)** 022501
[16] Shcheslavskiy V, Morozov P, Divochiy A, Vakhtomin Yu, Smirnov K and Becker W 2016 Ultrafast time measurements by time-correlated single photon counting coupled with superconducting single photon detector *Review of Scientific Instruments* **87** 053117
[17] Bell M, Sergeev A, Mitin V, Bird J, Verevkin A and Gol'tsman G 2007 *Phys. Rev. B* **76** 094521
[18] Kitaygorsky J et al 2005 *IEEE Trans. Appl. Supercond.* **15** 545
[19] Wolf S and Lowrey W H 1977 *Phys. Rev. Lett.* **39** 1038
[20] Kondo S 1992 Superconducting characteristics and the thermal stability of tungsten-based amorphous thin films Journal *of materials research* **7(04)** pp 853-860
[21] Vasil'ev D D, Malevannaya E I and Moiseev K M. 2015 Distribution of correlation between components around the substrate when applying WSi ultra-thin films from two sources by method of magnetron sputtering *Vacuum science and technics Materials of XXII scientific and technical Conf.* ed A S Bugaev (Moscow) pp 18-22
[22] Gol'tsman G N et al. 2003 Fabrication of nanostructured superconducting single-photon detectors *Applied Superconductivity, IEEE Transactions on*. **13** №2 pp192-195
[23] Shibata H, Shimizu K, Takesue H and Tokura Y 2013 Superconducting nanowire single-photon detector with ultralow dark count rate using cold optical filters *Applied Physics Express* **6(7)** 072801

**Acknowledgments**
The research was financially supported by the Ministry of Education and Science of the Russian Federation (Contract № 3.2655.2014/K).